\documentclass{article}
\pdfoutput=1
\usepackage{jheppub}
\usepackage[utf8]{inputenc}
\usepackage[english]{babel}
	\usepackage{feynmp}
	\def\bi{\begin{itemize}}
		\def\ei{\end{itemize}}
	\def\be{\begin{equation}}
	\def\ee{\end{equation}}
	\def\ba{\begin{align}}
	\def\ea{\end{align}}
	\def\bea{\begin{eqnarray}}
	\def\eea{\end{eqnarray}}
	\def\pd{\partial}
	\def\a{\alpha}
	\def\b{\beta}

	\def\m{\mu}
	\def\n{\nu}

	\def\l{\lambda}
	
	\def\r{\rho}
	\def\s{\sigma}
	\def\e{\epsilon}
	\def\wl{\widetilde{\l}}
	\def\wm{\widetilde{\m}}
	\def\da{\dot{\a}}

	\preprint{IFT-UAM/CSIC-16-043, FTUAM-16-16, UCM/FTI-314-16}
	\title{\boldmath  Unimodular Trees  versus Einstein Trees.}

	\author[a]{Enrique Alvarez,}
	\author[a]{Sergio Gonz\'alez-Mart\'{\i}n,}
	\author[b]{Carmelo P. Mart\'{\i}n.}

	\affiliation[a]{Departamento de F\'{\i}sica Te\'orica and Instituto de F\'{\i}sica Te\'orica, IFT-UAM/CSIC\\Universidad Aut\'onoma, 20849 Madrid, Spain}
	\affiliation[b]{Universidad Complutense de Madrid (UCM), Departamento de F\'{\i}sica Te\'orica I
		Facultad de Ciencias F\'{\i}sicas, Av. Complutense S/N (Ciudad Univ.), 28040 Madrid, Spain}

	\emailAdd{enrique.alvarez@uam.es}
	\emailAdd{sergio.gonzalez.martin@csic.es}
	\emailAdd{carmelop@fis.ucm.es}

	\abstract{The maximally helicity violating (MHV) tree level scattering amplitudes involving three, four or five gravitons are worked out in Unimodular Gravity. They are found to coincide with the corresponding amplitudes in General Relativity. This a remarkable result, insofar as both the propagators and the vertices are quite different in both theories.}

\begin{document}
	\maketitle
	    \unitlength = 1mm
	    \section{Introduction}
 Unimodular gravity is an interesting truncation of General Relativity , where the spacetime metric is restricted to be unimodular
 \be
 g\equiv
  \text{det}~g_{\m\n}=-1
 \ee
 It is convenient to implement the truncation through the (non invertible) map
 \be
 g_{\m\n}\longrightarrow |g|^{-1/n}~g_{\m\n}
 \ee
 The resulting theory is not Diff invariant anymore, but only TDiff invariant. Transverse diffeomorphisms are those such that their generator is transverse, that is
 \be
 \pd_\m \xi^\m=0
 \ee

 The ensuing action of Unimodular Gravity (cf. \cite{AlvarezGMHVM} for a recent review with references to previous literature),  reads

\be
S_{UG}\equiv \int d^n x~{\cal L}_{UG}\equiv -M_P^{n-2}\int |g|^{1/n}\left(R+{(n-1)(n-2)\over 4 n^2}{g^{\m\n}\nabla_\m g\nabla_\n g\over g^2}\right)
\ee
It can be easily shown using Bianchi identities that the classical equations of motion (EM) of Unimodular Gravity coincide with those of General Relativity with an arbitrary cosmological constant. The main difference at this level between both theories is that a constant value for the matter potential energy does not weight at all, which solves part of the cosmological constant problem (namely why the cosmological constant is not much bigger that observed). This property is preserved by quantum corrections \cite{Redux}.
\par
A natural question to ask at this stage is whether the S-matrix would be the same for Unimodular Gravity as for General Relativity. Although the S-matrix elements have been studied by several authors in the case of General Relativity \cite{Benincasa,Berends,Cachazo,Benincasa2,Ananth},  we are not aware of any results concerning the computation of  S-matrix elements in Unimodular Gravity. The propagators as well as the vertices are quite different in both theories, so that the answer to the question we asked at the beginning of this paragraph is not immediate.
\par

In the present paper we shall carry out the calculation of the maximally helicity violating  three, four and five graviton amplitudes at the tree-level and have found complete agreement between both theories, a fact that we find remarkable.


	    \section{Feynman rules}
	
	    The graviton propagator in  Unimodular Gravity (cf. Appendix A) reads

	    \begin{equation}
\begin{array}{l}
{P^\text{UG}_{\m\n,\r\s}=\dfrac{i}{2k^2}\left(\eta_{\m\s}\eta_{\n\r}+\eta_{\m\r}\eta_{\n\s}\right)-\dfrac{i}{k^2}\dfrac{\a^2n^2-n+2}{\a^2 n^2(n-2)}\eta_{\m\n}\eta_{\r\s}+\dfrac{2i}{n-2}\left(\dfrac{k_\r k_\s \eta_{\m\n}}{k^4}+\dfrac{k_\m k_\n \eta_{\r\s}}{k^4}\right)}\\[4pt]
{\phantom{P^\text{UG}_{\m\n,\r\s}=}
-\dfrac{2in}{n-2}\dfrac{k_{\m}k_{\n}k_{\r}k_{\s}}{k^6}}
\label{propagatorug}
 \end{array}
 \end{equation}	
for the gauge choice of \cite{AlvarezGMHVM}.

	Recall that the usual General Relativity graviton propagator in the de Donder gauge
	
	    \be P_{\m\n\r\s}^{\text{GR}}=\dfrac{i}{2k^2}\left(\eta_{\m\s}\eta_{\n\r}+\eta_{\m\r}\eta_{\n\s}-\dfrac{2}{n-2}\eta_{\m\n}\eta_{\r\s}\right)
	    \ee
	    has only simple poles at $k^2=0$. In the unimodular propagator, by contrast,
	    there appear double and triple poles in addition to the simple ones. This is a technical complication and the main reason why we can not, \textit{a priori}, apply some of the recent useful techniques \cite{Britto} to reduce the computation of the diagrams. In Appendix B we shall show that no gauge choice in Unimodular Gravity can yield a propagator of the form

	    \be P_{\m\n\r\s}=\dfrac{i}{2k^2}\left(\eta_{\m\s}\eta_{\n\r}+\eta_{\m\r}\eta_{\n\s}-f_1(k^2)\,\eta_{\m\n}\eta_{\r\s}\right)+f_2(k^2)
(k_\r k_\s \eta_{\m\n}+k_\m k_\n \eta_{\r\s})+f_3(k^2)\,k_{\m}k_{\n}k_{\r}k_{\s},
	    \ee
 $f_3(k^2)$ having no pole at $k^2=0$, if the Newtonian potential  is to be obtained in the nonrelativistic limit. Actually, we shall see that the triple pole term in (\ref{propagatorug}) is needed to retrieve the correct non-relativistic static limit.

	    Since we are going to focus on the three, four and five point amplitudes, we also need the three and four graviton vertex. These are obtained from the second and third order expansion of the Lagrangian around flat space (cf. Appendix C) and can be expressed in a condensed form, with a parameter $n$ that gives the General Relativity vertex for $n=2$ and the Unimodular Gravity one for $n=4$. With the convention of all incoming momenta the expression for the three-graviton vertex reads

	   \bea V^{\m\n,\r\s,\a\b}_{(p1,p2,p3)}&=&i \kappa ~{\cal S}\left\{ - \frac{\Bigl(2 + n\Bigr) (p_1.p_2) \eta^{\a\r} \eta^{\b\s} \eta^{\m\n}}{n^2}  -  \frac{(p_1.p_2) \eta^{\a\b} \eta^{\m\r} \eta^{\n\s}}{2 n} + \frac{\Bigl(2 + n\Bigr) (p_1.p_2) \eta^{\a\b} \eta^{\m\n} \eta^{\r\s}}{2 n^3}+\nonumber\right.\\
	   && + \frac{2 \eta^{\b\n} \eta^{\r\s} p_1^{m} p_2^{\a}}{n} + \tfrac{1}{2} \eta^{mr} \eta^{\n\s} p_1^{\a} p_2^{\b} -  \frac{\Bigl(2 + n\Bigr) \eta^{\m\n} \eta^{\r\s} p_1^{\a} p_2^{\b}}{2 n^2} - 2 \eta^{\b\s} \eta^{\n\r} p_1^{\a} p_2^{\m} -  \eta^{\a\n} \eta^{\b\s} p_1^{\r} p_2^{\m} +\nonumber \\
	   && + \frac{\eta^{\a\b} \eta^{\n\s} p_1^{\r} p_2^{\m}}{n} + \frac{2 \eta^{\b\m} \eta^{\r\s} p_1^{\a} p_2^{\n}}{n} -  \frac{2 \eta^{\a\b} \eta^{\r\s} p_1^{\m} p_2^{\n}}{n^2} + \frac{2 \eta^{\a\m} \eta^{\b\n} p_1^{\s} p_2^{\r}}{\n}+ (p_1.p_2) \eta^{\a\n} \eta^{\b\s} \eta^{\m\r}\Bigg\}\nonumber\\ \eea

The four-graviton vertex, in turn, is given by	
	
	   \bea
	  V^{\m\n,\r\s,\a\b,\eta\l}_{(p1,p2,p3,p4)}&=&i \kappa^2~  {\cal S}\Bigg\{ \frac{\Bigl(2 + n\Bigr) (p_3.p_4) g^{\m\n} g^{\r\s} g^{\a\b} g^{\eta\lambda}}{4 n^4} -  \frac{\Bigl(2 + n\Bigr) (p_3.p_4) g^{\m\r} g^{\a\b} g^{\eta\lambda} g^{\n\s}}{4 n^3}  +\nonumber\\
	  && + \frac{\Bigl(2 + n\Bigr) (p_3.p_4) g^{\m\eta} g^{\r\a} g^{\n\lambda} g^{\s\b}}{2 n^2} -  \frac{\Bigl(2 + n\Bigr) (p_3.p_4) g^{\m\n} g^{\r\eta} g^{\a\b} g^{\s\lambda}}{ n^3} +\nonumber\\
	  &&+ \frac{\Bigl(2 + n\Bigr) (p_3.p_4) g^{\m\r} g^{\a\b} g^{\eta\s} g^{\n\lambda}}{ n^2}-  \frac{(p_3.p_4) g^{\m\n} g^{\r\s} g^{\a\eta} g^{\b\lambda}}{4 n^2} + \frac{(p_3.p_4) g^{\m\n} g^{\r\eta} g^{\a\s} g^{\b\lambda}}{ n} +\nonumber\\
	  && +  g^{\m\eta} g^{\a\s} g^{\b\lambda} p_3^{\n} p_4^{\r} + \frac{\Bigl(2 + n\Bigr) g^{\m\r} g^{\a\b} g^{\eta\lambda} p_3^{\s} p_4^{\n}}{2 n^2} -  \tfrac{1}{2} g^{\m\r} g^{\a\eta} g^{\b\lambda} p_3^{\s} p_4^{\n} +\nonumber\\
	  &&+ \frac{\Bigl(2 + n\Bigr) g^{\m\a} g^{\eta\lambda} g^{\n\b} p_3^{\r} p_4^{\s}}{ n^2} + \frac{g^{\m\n} g^{\a\eta} g^{\b\lambda} p_3^{\r} p_4^{\s}}{2 n} -   g^{\m\a} g^{\eta\n} g^{\b\lambda} p_3^{\r} p_4^{\s} -  2\frac{g^{\m\a} g^{\r\b} g^{\eta\lambda} p_3^{\n} p_4^{\s}}{n}  -\nonumber\\
	  &&-  \frac{\Bigl(2 + n\Bigr) g^{\m\n} g^{\a\b} g^{\eta\lambda} p_3^{\r} p_4^{\s}}{2 n^3}-  2\frac{g^{\m\a} g^{\eta\s} g^{\n\b} p_3^{\r} p_4^{\lambda}}{n}+ 2\frac{g^{\m\n} g^{\r\a} g^{\eta\lambda} p_3^{\s} p_4^{\b}}{n^2} - \nonumber\\
	  &&-  2\frac{g^{\m\r} g^{\a\s} g^{\eta\lambda} p_3^{\n} p_4^{\b}}{n} + 2g^{\m\eta} g^{\r\lambda} g^{\a\n} p_3^{\s} p_4^{\b} -  2\frac{g^{\m\n} g^{\r\eta} g^{\a\lambda} p_3^{\s} p_4^{\b}}{n} -  2\frac{g^{\m\eta} g^{\r\a} g^{\n\lambda} p_3^{\s} p_4^{\b}}{n} + \nonumber\\
	  &&+ 2g^{\m\r} g^{\a\lambda} g^{\eta\s} p_3^{\n} p_4^{\b}-  2\frac{g^{\m\r} g^{\a\b} g^{\eta\s} p_3^{\n} p_4^{\lambda}}{n}+ 2\frac{g^{\m\n} g^{\r\eta} g^{\a\b} p_3^{\s} p_4^{\lambda}}{n^2}\nonumber\\
	  &&+ \frac{g^{\m\n} g^{\r\s} g^{\a\eta} p_3^{\lambda} p_4^{\b}}{2 n^2} -  \frac{g^{\m\n} g^{\r\eta} g^{\a\s} p_3^{\lambda} p_4^{\b}}{ n} +  g^{\m\r} g^{\a\n} g^{\eta\s} p_3^{\lambda} p_4^{\b} -  \frac{g^{\m\r} g^{\a\eta} g^{\n\s} p_3^{\lambda} p_4^{\b}}{2 n}  -\nonumber\\&&-  \frac{g^{\m\n} g^{\r\s} g^{\a\b} p_3^{\eta} p_4^{\lambda}}{ n^3} + \frac{g^{\m\r} g^{\a\b} g^{\n\s} p_3^{\eta} p_4^{\lambda}}{ n^2} -  2\frac{g^{\m\r} g^{\a\s} g^{\n\b} p_3^{\eta} p_4^{\lambda}}{n} + 2\frac{g^{\m\n} g^{\r\a} g^{\s\b} p_3^{\eta} p_4^{\lambda}}{n^2}  -\nonumber\\
	  &&-  \tfrac{1}{2} (p_3.p_4) g^{\m\eta} g^{\r\lambda} g^{\a\n} g^{\s\b}-  (p_3.p_4) g^{\m\r} g^{\a\n} g^{\eta\s} g^{\b\lambda} + \frac{(p_3.p_4) g^{\m\r} g^{\a\eta} g^{\n\s} g^{\b\lambda}}{4 n}\Bigg\}
	   	   \eea
	
	    Where {\cal S} is a shorthand for a double symmetrization, namely
	     \begin{enumerate}
	    	\item A summation over all momentum-index combinations ($p_1,{\m\n}$; $~p_2,\r\s$; $~;p_3,\a\b$; $~p_4,\eta\l$).
	    	\item A symmetrization of each pair on indices $\m\n$, $\r\s$, $\a\b$, $\eta\l$. \footnote{We have compared our vertices with those of \cite{Sannan,DeWitt}, in their notation, and in addition to the error pointed out in \cite{Sannan} in the four vertex, we claim that their last symbol is $2P_{12}$ instead of $4P_6$.}
	    	\end{enumerate}

	    	\section{Spinor helicity formalism for massless particles}
	    	
	       	Although we are no using the spinor helicity formalism explicitly, we can take advantage of some useful relationships that can be derived from it and will simplify greatly the calculations. \newline
	    	
	    	The four momentum $p^\m$ for an on-shell particle is written in terms of two commuting Weyl spinors as
	    	\be p_{\a\da}=\bar{\sigma}_{\m,\a\da}p^\m=\l_{\a}\wl_{\da}+\m_\a\wm_{\da} \ee
	
	    	 And in the case of a massless particle, the condition $\text{det}(p_{\a\da})=0$ implies
	    	
	    	 \be p_{\a\da}=\bar{\sigma}_{\m,\a\da}p^\m=\l_{\a}\wl_{\da} \ee
	    	
	    	On the other hand, the polarization tensor of the graviton can be written in terms of the gluon ones as
	    	
	    	\be \epsilon^-_{\m\n}=\epsilon^-_\m\epsilon^-_\n\longrightarrow\epsilon^-_{a\dot{a},b\dot{b}}=\epsilon^-_{a\dot{a}}\epsilon^-_{b\dot{b}}~~\text{and}~~\epsilon^+_{a\dot{a},b\dot{b}}=\epsilon^+_{a\dot{a}}\epsilon^+_{b\dot{b}}\ee
	    	
	    	The gluon polarization vector depends on the momentum of the given gluon and an arbitrary reference momentum $\e^-_{i,\m}\equiv\e^-_\m(p_i,r_i)$ where, following the conventions of \cite{Henn}, the gluon polarization spinors are given by
	    	
	    	\be \epsilon^-_{a\dot{a}}=\sqrt{2}\dfrac{\l_\a \widetilde{\m}_{\dot{a}}}{[\l\m]},\qquad \epsilon^+_{a\dot{a}}=-\sqrt{2}\dfrac{\widetilde{\l}_{\dot{\a}} {\m}_\a}{\langle\l\m\rangle}\ee
with $\mu$ and $\widetilde{\mu}$ the reference spinors which are related with the freedom to perform a  gauge transformation. Therefore, they can be chosen in such a way as to simplify the computations as much as possible: this is achieved by choosing the so called ``minimal gauge" --see \cite{Berends:1987cv}-- as displayed next.\newline

Altogether, this implies that, for any given particle

\be \e^+_i\cdot\e^-_i=-1,\qquad \e^+_i\cdot\e^+_i=\e^-_i\cdot\e^-_i =0,\qquad \e_i^{\pm}\cdot p_i=\e_i^{\pm}\cdot r_i=0\ee

Henceforth, with the appropriate choice of the reference spinors we get the following rules,

\begin{enumerate}
	\item For the four graviton amplitudes, by choosing $r_1=r_2=p_4$ and $r_3=r_4=p_1$ we get the extra relations:
	
	\bea
		&&\e_1^-\cdot p_4=0\\
			&&\e_2^-\cdot p_4=0\\
	&&\e_3^+\cdot p_1=0\\
		&&\e_4^+\cdot p_1=0\\
		&&\e_i^{\pm}\cdot \e_j^{\pm}=0~~\text{except for}~~ \e_2\cdot\e_3
	\eea
	
	\item For the five graviton amplitudes, we choose now $r_1=r_2=p_5$ and $r_3=r_4=r_5=p_1$ we get the extra relations:
	
	\bea
\qquad\qquad\qquad\qquad	&&\e_1^-\cdot p_5=0\\
	&&\e_2^-\cdot p_5=0\\
	&&\e_3^+\cdot p_1=0\\
	&&\e_4^+\cdot p_1=0\\
	&&\e_5^+\cdot p_1=0\\
	&&\e_i^{\pm}\cdot \e_j^{\pm}=0~~\text{except for}~~ \e_2^{\pm}\cdot\e_3^{\pm}~~\text{and}~~\e_2^{\pm}\cdot\e_4^{\pm}
	\eea

	\end{enumerate}

\section{Three graviton amplitudes}

The fact that Unimodular Gravity perturbatively expanded around Minkowski spacetime is Lorentz invariant and that the graviton polarizations are the same as in General Relativity leads, by repeating the standard analysis \cite{Elvang:2015rqa}, to the conclusion that the on-shell three-point amplitudes vanish on-shell for real momenta. Now, let us stress that little group scaling operates in Unimodular Gravity exactly in the manner as in General Relativity. Hence, it is plain that for conserved complex momenta the on-shell nonvanishing three-point amplitudes are the same in Unimodular Gravity as in General Relativity but, perhaps, for a global constant. By explicit computation of the corresponding Feynman diagrams we have found that the constant in question is same in both theories, as becomes the fact that the classical  Newton constant is indeed the same for both theories. Let us notice that the on-shell three-point functions for complex momenta are the elementary objects in the recursive construction of the amplitudes in theories like Yang-Mills and General Relativity with or without SUSY.

\section{Four graviton Tree Amplitudes}

Let us recall that our goal is to compute the tree diagrams both in Unimodular Gravity and General Relativity in order to see whether  there is any difference between both theories. This is relevant for the physical content of the theories because these amplitudes give us information on the tree level  S matrix.
\par

We shall focus on the maximally helicity violating (MHV) diagrams with three, four and five external gravitons because they are the simplest nontrivial ones.

	    	There are only three types of diagrams --which correspond to the well-known $s$, $t$ and $u$ channels, respectively-- that involve four external gravitons to be worked out explicitly.  The diagram that is a pure four vertex vanishes because no nonvanishing contribution to the amplitude diagram can be constructed out of two momenta entering the vertex and the four graviton polarizations satisfying the equations displayed in Section 3.
		   The $s$, $t$ and $u$-channel diagrams are shown in the next figures where all gravitons are outgoing.
	    	\newpage
	    	
	    	 \begin{figure}[h!]
	    	 	\centering
	    	 		  \begin{minipage}{0.45\linewidth}
	    	 		    		\centering
	    	 	\begin{fmffile}{4pointss}
	    	 		
	    	 		\begin{fmfgraph*}(80,50)	
	    	 			\fmfleft{i1,i2}
	    	 			\fmfright{e2,e1}
	    	 				\fmfset{curly_len}{1.5mm}
	    	 			\fmf{curly,tension=1,lab=\LARGE$p_1$}{i1,v1}
	    	 			\fmf{curly,tension=1,lab=\LARGE $p_2~$,l.side=left}{i2,v1}
	    	 			\fmf{curly,lab=\LARGE$q$,l.side=left}{v1,v2}
	    	 			\fmf{curly,tension=1,lab=\LARGE$p_3$}{e1,v2}
	    	 			\fmf{curly,tension=1,lab=\LARGE$p_4$,l.side=right}{e2,v2}
	    	 		\end{fmfgraph*}
	    	 		
	    	 	\end{fmffile}
	    	 	    	    	 	\caption{S channel}
	    	 		\end{minipage}
	    	    	\begin{minipage}{0.45\linewidth}
		\centering
	    	 	\centering
	    	 	\begin{fmffile}{4points2}

	    	 		\begin{fmfgraph*}(80,50)	
	    	 			\fmfleft{e2,i1}
	    	 			\fmfright{i2,e1}
	    	 				\fmfset{curly_len}{1.5mm}
	    	 			\fmf{curly,tension=1,lab=\LARGE$p_1$,l.side=left}{i1,v1}
	    	 			\fmf{curly,tension=1,lab=\LARGE$p_4~~~$}{i2,v2}
	    	 			\fmf{curly,lab=\LARGE$q$,l.side=left}{v1,v2}
	    	 			\fmf{curly,tension=1,lab=\LARGE$p_3$,label.dist=2.5mm}{e1,v1}
	    	 			\fmf{curly,tension=1,lab= \LARGE$~~ p_2$,l.side=right}{e2,v2}
	    	 			
	    	 		\end{fmfgraph*}
	    	 		   	 		
	    	 	\end{fmffile}
	    	 	\caption{T channel}
	    	 		\end{minipage}
	    	 \end{figure}
	    	 	    	 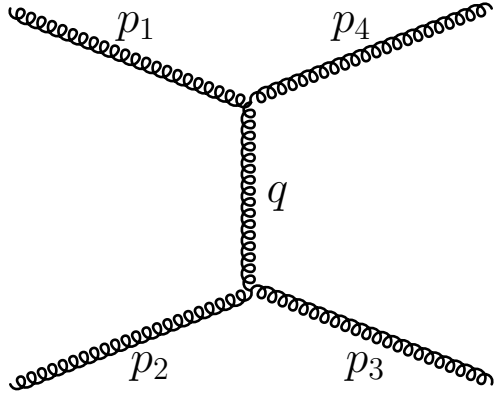
\begin{figure}[h!]
	    	 	\centering
	    	 	\begin{fmffile}{4pointsu}

	    	 		\begin{fmfgraph*}(80,50)
	    	 			\fmfleft{e1,i1}
	    	 			\fmfright{i2,e2}
	    	 				\fmfset{curly_len}{1.5mm}
	    	 			\fmf{curly,tension=1,lab=\LARGE$p_1$,l.side=left}{i1,v1}
	    	 			\fmf{curly,tension=1,lab=\LARGE$p_3$,label.angle=140}{i2,v2}
	    	 			\fmf{curly,lab=\LARGE$q$,l.side=left}{v1,v2}
	    	 			\fmf{curly,tension=1,lab=\LARGE$~~p_2$}{e1,v2}
	    	 	
	    	 			\fmf{curly,tension=1,lab=\LARGE$p_4~~$,l.side=right}{e2,v1}

	    	 		\end{fmfgraph*}
	    	   	
	    	   \end{fmffile}
	    	   \caption{U channel}
		    	\end{figure}
	    		    The explicit result is   	
		 \bea
	   {\cal A}_s(1^-2^-;3^+4^+)&=&\e_1^{-\m_1}\e_1^{-\n_1}\e_2^{-\m_2}\e_2^{-\n_2}V^{\m_1\n_1,\m_2\n_2,\a\b}_{(p1,p2,q)}P_{\a,\b,\r,\s}V^{\r\s,\m_3\n_3,\m_4\n_4}_{(p,p3,p4)}\e_3^{-\m_3}\e_3^{-\n_3}\e_4^{-\m_4}\e_4^{-\n_4}\nonumber\\&=&- \frac{i \kappa^2 (\e_1.p_2)^2 (\e_2.\e_3)^2 (\e_4.p_2)^2}{s^2}=i\dfrac{\kappa^2}{4}\dfrac{\langle 12\rangle ^5[34]^2}{[12]\langle 23\rangle ^2\langle 14\rangle ^2}\\
	    {\cal A}_t(1^-2^-;3^+4^+)&=&\e_1^{-\m_1}\e_1^{-\n_1}\e_3^{+\m_3}\e_3^{+\n_3}V^{\m_1\n_1,\m_3\n_3,\a\b}_{(p1,p3,q)}P_{\a,\b,\r,\s}V^{\r\s,\m_2\n_2,\m_4\n_4}_{(p,p2,p4)}\e_2^{-\m_2}\e_2^{-\n_2}\e_4^{+\m_4}\e_4^{+\n_4}\nonumber\\&=&0\\
	     {\cal A}_u(1^-2^-;3^+4^+)&=&\e_1^{-\m_1}\e_1^{-\n_1}\e_4^{+\m_4}\e_4^{+\n_4}V^{\m_1\n_1,\m_4\n_4,\a\b}_{(p1,p4,q)}P_{\a,\b,\r,\s}V^{\r\s,\m_2\n_2,\m_3\n_3}_{(p,p2,p3)}\e_2^{-\m_2}\e_2^{-\n_2}\e_3^{+\m_3}\e_3^{+\n_3}\nonumber\\&=&
	       \frac{i \kappa^2 (\e_1.p_2)^2 (\e_2.\e_3)^2 (\e_4.p_2)^2}{u^2}=i\dfrac{\kappa^2}{4}\dfrac{\langle 12\rangle ^8[24]^2}{\langle 13\rangle ^3[31]\langle 23\rangle ^2\langle 14\rangle ^2}
	 \eea
	
	 where as usual, $s=p_1+p_2$ and $u=p_1+p_3$.
	
		 These amplitudes are \textit{diagram to diagram} exactly the same that the ones for General Relativity. The complete amplitude is therefore
				 \be{\cal A}(1^-2^-;3^+4^+)=i\dfrac{\kappa^2}{4}\dfrac{\langle 12\rangle ^8[12]}{\langle 12\rangle\langle 13\rangle\langle 14\rangle\langle 23\rangle\langle 24\rangle\langle 34\rangle ^2}\ee
		 in agreement with the result presented for General Relativity in \cite{Cachazo}.
\section{ Five point diagrams}

	      When computing the diagrams with five external gravitons there are three sets of diagrams. The one that is purely a 5-vertex vanishes identically. Indeed, no nonvanishing contribution to the amplitude diagram  can be built from two momenta entering the vertex and the five graviton polarizations introduced in Section 3. Let us consider the others in turn
\subsection{Three vertices}
	
\begin{figure}[h!]
	      	\centering
	      	\begin{fmffile}{5point}
	      		
	      		\begin{fmfgraph*}(80,50)	
	      			\fmfleft{i1,i2}
	      			\fmfright{e2,e1}
	      			\fmftop{o1}
	      			\fmfset{curly_len}{1.5mm}
	      			\fmf{curly,tension=1,l.side=left,lab=\Large$p_1$}{i1,v1}
	      			\fmf{curly,tension=1,l.side=right,lab=\Large$p_2~$}{i2,v1}
	      			\fmf{curly,lab=\Large$q$,l.side=left}{v1,v3}
	      			\fmf{curly,lab=\Large$k$,l.side=left}{v3,v2}
	      			\fmf{curly,tension=1,lab=\Large$p_4$}{e1,v2}
	      			\fmf{curly,tension=1,lab=\Large$~~p_5$,l.side=right}{e2,v2}
	      			\fmf{curly,tension=0,lab=\Large$p_3~~$}{o1,v3}

	      		\end{fmfgraph*}	      	
	      	\end{fmffile}
	     	      	\caption{ ${\cal A}(1^-,2^-;3^+;4^+,5^+)$}
	     	      	 \label{3vertex}
	      \end{figure}
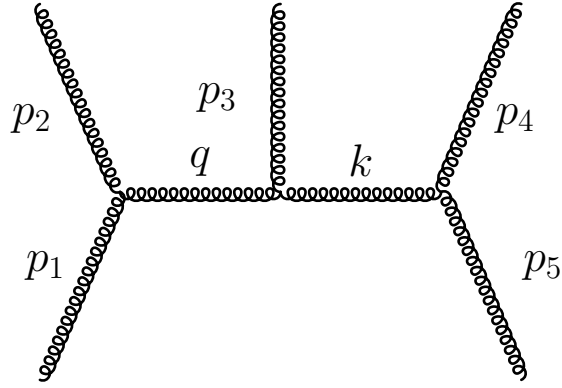

	      There are 15 different diagrams that involve three three-vertex of the type shown in Figure \ref{3vertex}; this we shall denote by ${\cal A}(1^-,2^-;3^+;4^+,5^+)$, the others will be analogously represented by using the obvious notation.

	       Let us write this one as example; the full set of amplitudes can be found in the Appendix D.

	       \bea
	       {\cal A}(1^-,2^-;3^+;4^+,5^+)&&=- \frac{i \kappa^3 (\e_1.p_2)^2 (\e_2.\e_4)^2 (\e_3.p_2)^2 (\e_5.p_2)^2}{(p_1+p_2)^2 (p_4+p_5)^2} -  \frac{i \kappa^3 (\e_1.p_2)^2 (\e_2.\e_3)^2 (\e_4.p_2)^2 (\e_5.p_3)^2}{(p_1+p_2)^2 (p_4+p_5)^2}-\nonumber\\
	       -&&  \frac{i \kappa^3 (\e_1.p_2)^2 (\e_2.\e_3)^2 (\e_4.p_3)^2 (\e_5.p_2)^2}{(p_1+p_2)^2 (p_4+p_5)^2} -  \frac{i \kappa^3 (\e_1.p_2)^2 (\e_2.\e_4)^2 (\e_3.p_2)^2 (\e_5.p_3)^2}{(p_1+p_2)^2 (p_4+p_5)^2}-\nonumber\\-&&  \frac{2i \kappa^3 (\e_1.p_2)^2 (\e_2.\e_4)^2 (\e_3.p_2)^2 (\e_5.p_2) (\e_5.p_3)}{(p_1+p_2)^2 (p_4+p_5)^2} +\nonumber\\
	       +&& \frac{2i \kappa^3 (\e_1.p_2)^2 (\e_2.\e_3) (\e_2.\e_4) (\e_3.p_2) (\e_4.p_2) (\e_5.p_2) (\e_5.p_3)}{(p_1+p_2)^2 (p_4+p_5)^2} -\nonumber\\
	       +&& \frac{2i \kappa^3 (\e_1.p_2)^2 (\e_2.\e_3) (\e_2.\e_4) (\e_3.p_2) (\e_4.p_3) (\e_5.p_2) (\e_5.p_3)}{(p_1+p_2)^2 (p_4+p_5)^2} +\nonumber\\
	       +&& \frac{2i \kappa^3 (\e_1.p_2)^2 (\e_2.\e_3)^2 (\e_4.p_2) (\e_4.p_3) (\e_5.p_2) (\e_5.p_3)}{(p_1+p_2)^2 (p_4+p_5)^2}  +\nonumber\\
	       +&& \frac{2i \kappa^3 (\e_1.p_2)^2 (\e_2.\e_3) (\e_2.\e_4) (\e_3.p_2) (\e_4.p_2) (\e_5.p_3)^2}{(p_1+p_2)^2 (p_4+p_5)^2} \nonumber\\
	       -&&  \frac{2i \kappa^3 (\e_1.p_2)^2 (\e_2.\e_3) (\e_2.\e_4) (\e_3.p_2) (\e_4.p_3) (\e_5.p_2)^2}{(p_1+p_2)^2 (p_4+p_5)^2}
	       \eea
	
\subsection{The four vertex}

\begin{figure}[h!]
	     	     	\centering
	     	     	\begin{fmffile}{5point4}    	
	     	     		\begin{fmfgraph*}(80,50)
	     	     			\fmfleft{i1,i2}
	     	     			\fmfright{e3,e2,e1}
	     	     				\fmfset{curly_len}{1.5mm}
	     	     			\fmf{curly,tension=1,lab=\Large$~p_1$}{i1,v1}
	     	     			\fmf{curly,tension=1,lab=\Large$p_2~~$}{i2,v1}
	     	     			\fmf{curly,lab=\Large$q$,l.side=left}{v1,v2}
	     	     			\fmf{curly,tension=1,lab=\Large$p_3~$}{e1,v2}
	     	     			\fmf{curly,tension=1,lab=\Large$~p_4$,l.side=right}{e2,v2}
	     	     			\fmf{curly,tension=1,lab=\Large$p_5$}{e3,v2}
	     	     		\end{fmfgraph*}
	     	     	\end{fmffile}
	     	     	\caption{ ${\cal A}(1^-,2^-;3^+,4^+,5^+)$}
	     	     	\label{4vertex}
	     	     \end{figure}
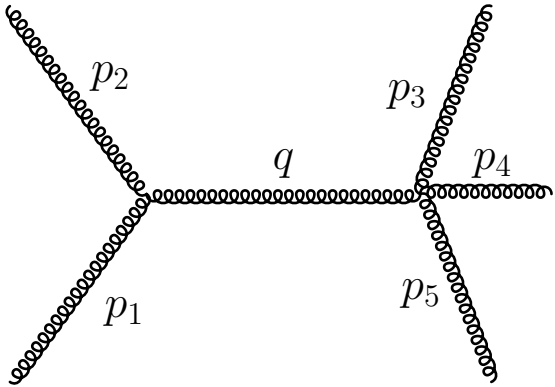

	     The rest of the  \textit{a priori} non-vanishing diagrams are those that involve one three vertex and one four vertex ${\cal A}(1^-,2^-;3^+,4^+,5^+)$ as shown in Figure \ref{4vertex}
	
   Explicit computation shows that all the 10 different diagrams do vanish.
 
\section{Conclusions}
It has been shown that the MHV three, four and five graviton tree amplitudes give the same contribution both in General Relativity and Unimodular Gravity. This result holds for each diagram independently and not only for the whole amplitude.
     Therefore we can conclude that, at least at the tree-level and for three, four or five external legs, the MHV contribution to the S matrix for pure Unimodular Gravity without coupling to other fields is the same in both theories.

     A remarkable fact  is that all the terms that involve the double and triple poles in the propagator of Unimodular Gravity \eqref{propagatorug} do not contribute to any diagram we have computed in pure Unimodular Gravity. We have explicitly checked  this by introducing an arbitrary coefficient in front of each piece and then verify that the final result is independent of the arbitrary coefficient we have introduced.
     That the contributions coming from those higher order poles go away is not trivial and we did not find any reason to expect it before computing the diagrams. Indeed, on the one hand, the triple pole summand in the propagator is needed to recover the Newtonian potential --see Appendix B- and, on the other hand, in Unimodular Gravity, one obtains the following nonzero result
     \be
  k_{\a} k_{\b} V^{\m\n,\r\s,\a\b}_{(p,q,k)}\, \e_{1\,\m\n}(p)\,\e_{2\,\r\s}(q) = i\kappa\,(p\cdot q)\,(p\cdot \e_{2})(q\cdot\e_{1})\, (\e_{1}\cdot\e_{2})
     \ee
     when $k=-p-q$ is off-shell and the polarizations with well-defined helicity $\e_{1\,\m\n}(p)=\e_{1\,\m}(p)\e_{1\,\n}(p)$ and $\e_{2\,\r\s}(q)=\e_{2\,\r}(q)\e_{2\,\s}(q)$   are arbitrary. This is in contrast with the fact that the computation of the corresponding object in General Relativity  yields a vanishing result as a consequence of invariance under the full Diffeomorphism group.
     \par
      As a straightforward consequence, and since the BFCW recursion relations \cite{Britto} can be applied to the diagrams of General Relativity \cite{Benincasa}, our results  suggest that BFCW (or a similar recurrence) can be  applied to Unimodular Gravity as well.
       This would be remarkable because of the existence of higher order poles in the propagator.
      Work on these issues is ongoing, and we expect to report on them soon.

\section{Acknowledgments}
      We are grateful for useful correspondence with W.T. Giele, L. Dixon and M. Spradlin. We also acknowledge useful discussions with Paolo Benincasa. This work has been partially supported by the European Union FP7 ITN INVISIBLES (Marie Curie Actions, PITN- GA-2011- 289442)and (HPRN-CT-200-00148); COST action MP1405 (Quantum Structure of Spacetime), COST action MP1210 (The String Theory Universe) as well as by FPA2012-31880 (MICINN, Spain)), FPA2014-54154-P (MICINN, Spain), and S2009ESP-1473 (CA Madrid).  This project has received funding from the European Union' s Horizon 2020 research and innovation programme under the Marie Sklodowska-Curie grant agreement No 690575. This project has also received funding from the European Union' s Horizon 2020 research and innovation programme under the Marie Sklodowska-Curie grant agreement No 674896.
      The authors acknowledge the support of the Spanish MINECO {\em Centro de Excelencia Severo Ochoa} Programme under grant SEV-2012-0249.
      
      Each  tree-level diagram workout in this paper has been computed in two independent ways, one using the computer algebra systems FORM \cite{Kuipers:2012rf} and the other Mathematica's xAct \cite{Martingarcia} package.

      \newpage
     \appendix
\section{Feynman rules}
 	
In order to  to obtain the Feynman rules for Unimodular Gravity, let us start from the action
	
	    \be\label{UG}
	    S_{UG}= - \dfrac{2}{\kappa^2}\int d^n x g^{1/n}\left(R+{(n-1)(n-2)\over 4 n^2}{\nabla_\m g\nabla^\m g\over g^2}\right)\ee
	
	    with $\kappa^2=32\pi G$.
	    	
	    The propagator is obtained by inverting the second order expansion of the Lagrangian around flat space-time- once properly gauge-fixed- presented in \cite{AlvarezGMHVM}. This reads, 	      \begin{align}
	      \nonumber {\cal L}&=\frac{1}{4}h^{\m\n}\partial^2 h_{\m\n}-\frac{1}{4 n}h \partial^2 h  +\left(-f\partial^2 f +\frac{\alpha}{2}f\partial^2 h+\frac{\alpha}{2}h\partial^2 f\right)-\\
	    &-\frac{1}{2}\left(\partial_{\m}c'^{\;(0,0)}\partial^{\m}c'^{\;(0,0)}+2\left(\partial_{\n}h^{\n}_{\m}-\frac{1}{n}\partial_{\m}h\right)\partial^{\m}c'^{\;(0,0)}\right)
	      \end{align}
	
	      Writing now the action as
	
	    \begin{align}\label{form_operator}
	    S=\int d^{n}x\; \Psi^{A}F_{AB}\Psi^{B}
	    \end{align}
	
	    where
	
	    \begin{align}
	    F_{AB}=G_{AB}\partial^2+ J^{\m\n}_{AB}\partial_{\m}\partial_{\n}
	    \end{align}
	
	      \begin{align}
	      \Psi^{A}=\begin{pmatrix}
	      h^{\m\n}\\
	      f\\
	      c'\\
	      \end{pmatrix}
	  \end{align}

	    and the different matrices involved read
	    \begin{align}
	    G_{AB}=\begin{pmatrix}
	    -\frac{1}{4}\left(\frac{1}{4}{\cal K}_{\m\n\r\s}^{\a\b}-{\cal P}_{\m\n\r\s}^{\a\b}\right)g_{\a\b} \quad&\frac{\a}{2}g_{\m\n}& -\frac{1}{8}g_{\m\n}\\
	    \frac{\a}{2} g_{\r\s}  & -1 &0\\
	    -\frac{1}{8}g_{\r\s}&0&\frac{1}{2}
	    \end{pmatrix}\label{matrix}
	    \end{align}
	
	    \begin{align}
	    J^{\a\b}_{AB}=\begin{pmatrix}
	    0&0&\frac{1}{4}\left(g^{\a}_{\m}g^{\b}_{\n}+g^{\a}_{\n}g^{\b}_{\m}\right)\\
	    0&0&0\\
	    \frac{1}{4}\left(g^{\a}_{\r}g^{\b}_{\s}+g^{\a}_{\s}g^{\b}_{\r}\right)&0&0
	    \end{pmatrix}
	    \end{align}
	
	    We have introduced the tensors
	    \begin{align}
	    {\cal P}_{\m\n\r\s}^{\a\b}&=\frac{1}{4}\left(g_{\m\r}\delta_{\n}^{(\a}\delta_{\s}^{\b)}+g_{\m\s}\delta^{(\a}_{\n}\delta^{\b)}_{\r}+g_{\n\r}\delta^{(\a}_{\m}\delta^{\b)}_{\s}+g_{\n\s}\delta^{(\a}_{\m}\delta^{\b)}_{\r}\right)\\
	    {\cal K}_{\m\n\r\s}^{\a\b}&=\frac{1}{2}\left(g_{\m\n}\delta^{(\a}_{\r}\delta^{\b)}_{\s}+g_{\r\s}\delta^{(\a}_{\m}\delta^{\b)}_{\n}\right)
	    \end{align}
\section{The Unimodular Gravity free propagator and Newton's Law}
 In Unimodular Gravity the graviton field $h_{\m\n}$ couples to the traceless part, $\hat{T}^{\m\n}$, of the Energy-momentum tensor {\it \`a la} Rosenfeld or, what is the same, the traceless part of the graviton field, $\hat{h}_{\m\n}$, couples to
the the Energy-momentum tensor defined {\it \`a la} Rosenfeld. Indeed,
\be -\frac{\kappa}{2}\int d^{4}x\;h_{\m\n}\,\hat{T}^{\m\n}=-\frac{\kappa}{2}\int d^{4}x\;\hat{h}_{\m\n}T^{\m\n},
\label{mattercoupling}
\ee
where
\be    \hat{T}^{\m\n}=T^{\m\n}-\frac{1}{4}\,T\,\eta^{\m\n}\quad\text{and}\quad  \hat{h}_{\m\n}=h_{\m\n}-\frac{1}{4}\,h\,\eta_{\m\n},
\ee
with $T=T^{\m}_{\m}$ and $h=h^{\m}_{\m}$.

 The Newtonian potential can be obtained \cite{Donoghue:1994dn} from the  tree-level one-graviton exchange, with transfer momentum $k_\m$, between two very massive scalar particles by taking  the so-called static limit: $k_\m=(0,\vec{k})_\m$. Let
$A_{12}$ denote the amplitude for the one-graviton exchange between two scalar particles with masses $M_1$ and $M_2$, respectively. In Unimodular Gravity --see equation (\ref{mattercoupling})-- we have
\be
A_{12}=-i\frac{\kappa^2}{4}T_{\m\n}^{1}(p_1,p'_1)\langle \hat{h}^{\m\n}(k) \hat{h}^{\r\s}(-k)\rangle T_{\r\s}^{2}(p_2,p'_2),
\ee
where $k=p_1-p'_1=p'_2-p_2$ and $p^2_i={p'}^2_i=M_i^2$, $i=1,2$. In the previous equation $\langle\hat{h}^{\m\n}(k) \hat{h}^{\r\s}(-k)\rangle$ denotes the free two-point function of the traceless graviton field and  $T_{\m\n}^{i}(p_i,p'_i)$, $i=1,2$, denote the lowest order contribution to the
on-shell matrix elements of the energy-momentum tensor between (on-shell) states with momentum $p_i$ and $p'_i$, $i=1,2$, respectively:
\be T_{\m\n}^{i}(p_i,p'_i)=p_{i\,\m}p'_{i\,\n}+p_{i\,\n}p'_{i\,\m}+\frac{1}{2}\,k^2\eta_{\mu\n}.
\ee
Now, for very massive particles and  for $k_\mu=(0,\vec k)$, we have
\be  \frac{1}{2M_i}T_{\m\n}^{i}(p_i,p'_i)= M_{i}\,\eta_{\m 0}\eta_{\n 0},\quad i=1,2
\ee
so that, in the static limit, one gets
\be
\frac{1}{2M_1\,2M_2}A_{12}=-i\frac{\kappa^2}{4}\,M_1\,M_2\,\langle\hat{h}^{00}(k) \hat{h}^{00}(-k) \rangle,
\label{nonrelati}
\ee
with $k_\m=(0,\vec{k})_\m$. It is the right hand side of the previous equation which must be equal to the Newtonian potential in Fourier space $V_{Nw}(\vec{k})$, where
\be
V_{Nw}(\vec{k})=-\frac{\kappa^2}{8}\, \frac{M_1 M_2}{\vec{k}^2}.
\label{Newtpotential}
\ee

Let us make the following ansatz for the free graviton two-point function, $<h_{\m\n}(k)h_{\r\s}(-k)>$, in Unimodular Gravity:
\bea
 \langle h_{\m\n}(k)h_{\r\s}(-k)\rangle=&&\frac{i}{2k^2}\left(\eta_{\m\s}\eta_{\n\r}+\eta_{\m\r}\eta_{\n\s}-\eta_{\m\n}\eta_{\r\s}\right)-i\frac{a(k^2)}{2k^2}\eta_{\m\n}\eta_{\r\s}+i\frac{b(k^2)}{(k^2)^2}\left(k_\r k_\s \eta_{\m\n}+k_\m k_\n \eta_{\r\s}\right)\nonumber\\&&+i\frac{c(k^2)}{(k^2)^3}\,k_{\m}k_{\n}k_{\r}k_{\s},
\label{propansatz}
 \eea
 where $a(k^2)$, $b(k^2)$ and $c(k^2)$ are arbitrary real functions. This ansatz is the most general expression consistent with Lorentz covariance, boson symmetry, the fact that $h_{\m\n}$ is a symmetric tensor and that when one replaces in the free two-point function the tensor
\be
\frac{1}{2}\left(\eta_{\m\s}\eta_{\n\r}+\eta_{\m\r}\eta_{\n\s}-\eta_{\m\n}\eta_{\r\s}\right)
\ee
with the following sum over polarizations,
\be
\sum_{\lambda=\pm 2}\e^{(\lambda)}_{\m\n}\e^{(-\lambda)}_{\r\s}
\ee
only a simple pole factor $1/k^2$ multiplies this sum, as befits the unitarity and the fact that the classical action of the theory is quadratic in the derivatives.

From equation (\ref{propansatz}), one obtains after a little algebra
\bea
\langle \hat{h}_{\m\n}(k)\hat{h}_{\r\s}(-k)\rangle=&& \frac{i}{2k^2}\left(\eta_{\m\s}\eta_{\n\r}+\eta_{\m\r}\eta_{\n\s}+\left(-\frac{1}{2}+\frac{c(k^2)}{8}\right)\eta_{\m\n}\eta_{\r\s}\right)\nonumber\\&&+i
\frac{c(k^2)}{4(k^2)^2}\left(k_\r k_\s \eta_{\m\n}+k_\m k_\n \eta_{\r\s}\right)+i\frac{c(k^2)}{(k^2)^3}\,k_{\m}k_{\n}k_{\r}k_{\s}.
\eea
Substituting the previous result in equation (\ref{nonrelati}) --recall that $k_\m=(0,\vec{k})_\m$, one gets
\be
-i\frac{\kappa^2}{4}\,M_1\,M_2\,\langle\hat{h}^{00}(k) \hat{h}^{00}(-k)\rangle =-\frac{\kappa^2}{8}\,M_1\,M_2\,\left(\frac{3}{2}+\frac{c(-\vec{k}^2)}{8}\right)\,\frac{1}{\vec{k}^2}.
\ee
This expression will match the Newtonian potential in (\ref{Newtpotential}) if, and only if, $c(-\vec{k}^2)=-4$, which, by Lorentz invariance, leads to
\be
c(k^2)=-4,
\ee
whatever the value of $k_\mu$. In summary, we need a triple pole in the $k_{\m}k_{\n}k_{\r}k_{\s}$ contribution to two-point function in (\ref{propansatz}) to get the Newtonian potential right. This is what actually happens when one
works out the propagator of Unimodular Gravity by using the BRST technique explained in  \cite{AlvarezGMHVM}. Notice that the propagator in \eqref{propagatorug} yields the Newtonian potential since the coefficient multiplying
the contribution
\be
\dfrac{k_{\m}k_{\n}k_{\r}k_{\s}}{k^6}
\ee
is $-4$, at $n=4$.

     \section{Expansion of the Unimodular Gravity Lagrangian}

  Starting from the action \eqref{UG} we perform a background field expansion of the metric around Minkowski $g_{\m\n}=\eta_{\m\n}+\kappa h_{\m\n}$ so it can be written as

     \be S_{UG}=-\dfrac{2}{\kappa^2}\int d^n x\left({\cal L}_0+\kappa{\cal L}_1+\kappa^2{\cal L}_2+\kappa^3{\cal L}_3+...\right)
     \ee

     Keeping $n$ free at this point it is worth to notice that this expansion will reduce to the General Relativity one taking $n=2$.

      As we are expanding Minkowski the first two terms vanish and the others read\footnote{For the General Relativity expansion we find a discrepancy with the expansion given in \cite{Goroff} for the third order lagrangian; the term proportional to $h\nabla^\m h\nabla_\m h$ has the opposite sign.}

     \begin{align}
     {\cal L}_{2}&=\frac{1}{4}h^{\m\n}\partial^2 h_{\m\n}-\frac{n+2}{4 n^{2}}h \partial^2 h  +\frac{1}{2} \left(\partial_{\m}h^{\m\a}\right)\left(\partial_{\n}h^{\n}_{\a}\right) -\frac{1}{n}\left(\partial_{\m}h\right)\left(\partial_{\n}h^{\m\n} \right)\nonumber\\
     {\cal L}_{3}&=-\dfrac{3}{4}h^{\m\n}\partial_\m h^{\a\b}\partial_\n h_{\a\b}+\dfrac{(3n-2)h^{\m\n}\partial_\m h\partial_\n h}{4n^2}-h^{\m\n}\partial_\n h\partial_\b h_\m^\b-h^{\m\n}\partial_\n h_\m^\b\partial_\b h-h_\m^\n h^{\m\b}\partial_\n\partial_\b h+\nonumber\\
     &+\dfrac{1}{n}h h^{\m\n}\partial_\n\partial_\m h+h_\m^\n h^{\m\b}\partial_\n\partial_\a h_\b^\a-\dfrac{1}{n}h h^{\m\n}\partial_\n\partial_\b h_\m^\b+\dfrac{(3n-2)h^{\m\n}\partial_\a h\partial^\a h_{\m\n}}{2n^2}-\dfrac{(3n-2)h\partial_\m h\partial^\m h}{4n^3}+\nonumber\\
     &+h^{\m\n}\partial_\a h_\m^\a\partial_\b h_\n^\b+2h^{\m\n}\partial_\n h_\m^\a\partial_\b h_\a^\b-\dfrac{1}{n}h\partial_\m h^{\m\n}\partial_\b h_\n^\b-h^{\m\n}\partial^\a h_{\m\n}\partial_\b h_\a^\b+\dfrac{1}{n}h\partial^\m h\partial_\b h_\m^\b+\nonumber\\
     &+h^{\m\n}h^{\a\b}\partial_\b\partial_\n h_{\m\a}-h^{\m\n}h^{\a\b}\partial_\b\partial_\a h_{\m\n}+h_\m^\n h^{\m\b}\partial_\a\partial_\n h_\b^\a-\dfrac{1}{n}hh^{\m\n}\partial_\a\partial_\n h_\m^\a-\dfrac{1}{2n}h_{\m\n}h^{\m\n}\partial_\a\partial_\b h^{\b\a}+\nonumber \\
     &+\dfrac{1}{2n^2}h^2\partial_\a\partial_\b h^{\b\a}-h_\m^\n h^{\m\a}\partial^2 h_{\a\n}+\dfrac{1}{n}h h^{\m\n}\partial^2 h_{\m\n}+\dfrac{1}{2n}h_{\m\n}h^{\m\n}\partial^2 h-\dfrac{1}{2n^2}h^2\partial^2 h+h^{\m\n}\partial_\n h_{\a\b}\partial^\b h_\m^\a+\nonumber \\
     &+\dfrac{1}{2}h^{\m\n}\partial_\a h_{\n\b}\partial^\b h_{\m}^\a-\dfrac{3}{2}h^{\m\n}\partial_\a h_{\n\b}\partial^\a h_\m^\b-\dfrac{1}{2n}h\partial_\m h_{\a\b}\partial^\b h^{\a\m}+\dfrac{3}{4n}h\partial_\m h_{\a\b}\partial^\m h^{\a\b}\nonumber\\
     \end{align}

     Integrating by parts and discarding total derivatives the cubic term can be written as
     \begin{align}
     {\cal L}_{3}&=\dfrac{n+2}{4n^3}h\partial_\m h\partial^\m h-\dfrac{n+2}{2n^2}h^{\m\n}\partial_\a h\partial^\a h_{\m\n}-\dfrac{1}{4n}h\partial_\a h^{\m\n}\partial^\a h_{\m\n}-\dfrac{1}{n^2}h\partial_\m h\partial_\n h^{\m\n}-\dfrac{n+2}{4n^2} h^{\m\n}\partial_\m h\partial_\n h+\nonumber\\
     &+\dfrac{1}{4}h^{\m\n}\partial_\m h^{\a\b}\partial_\n h_{\a\b}+\dfrac{1}{n}h^{\a\b}\partial^\m h_{\a\b}\partial^\n h_{\m\n}+\dfrac{1}{n}h^{\m\n}\partial_\n h\partial_\b h^\b_{\m}+\dfrac{1}{2n}h\partial^\a h^{\m\n}\partial_\n h_{\a\m}+\dfrac{1}{n}h^{\m\n}\partial_\b h\partial_\n h^\b_{\m}+\nonumber\\
     &+ \dfrac{1}{2}h^{\m\n}\partial_\a h_{\n\b}\partial^\a h^\b_\m-\dfrac{1}{2}h^{\m\n}\partial^\a h_{\n\b}\partial^\b h_{\a\m}-h^{\a\b}\partial_\n h_{\m\a}\partial_\b h^{\m\n}\end{align}

  \section{The full set of five-graviton tree diagrams}
  	
	       \bea
	       {\cal A}(1^-,2^-;4^+;3^+,5^+)&&=\frac{i \kappa^3 (\e_1.p_2)^2 (\e_2.\e_4)^2 (\e_3.p_5)^2 (\e_5.p_3) (\e_5.p_4)}{(p_1+p_2)^2 (p_3+p_5)^2} -  \frac{i \kappa^3 (\e_1.p_2)^2 (\e_2.\e_3)^2 (\e_4.p_2)^2 (\e_5.p_3)^2}{(p_1+p_2)^2 (p_3+p_5)^2}-\nonumber\\
	       -&&  \frac{2i \kappa^3 (\e_1.p_2)^2 (\e_2.\e_3) (\e_2.\e_4) (\e_3.p_5) (\e_4.p_2) (\e_5.p_2) (\e_5.p_3)}{(p_1+p_2)^2 (p_3+p_5)^2} - \nonumber\\-&&  \frac{i \kappa^3 (\e_1.p_2)^2 (\e_2.\e_4)^2 (\e_3.p_4) (\e_3.p_5) (\e_5.p_3)^2}{(p_1+p_2)^2 (p_3+p_5)^2} - \frac{i \kappa^3 (\e_1.p_2)^2 (\e_2.\e_4)^2 (\e_3.p_4)^2 (\e_5.p_3)^2}{(p_1+p_2)^2 (p_3+p_5)^2}  -\nonumber\\
	       -&&  \frac{2i \kappa^3 (\e_1.p_2)^2 (\e_2.\e_3) (\e_2.\e_4) (\e_3.p_5) (\e_4.p_2) (\e_5.p_3)^2}{(p_1+p_2)^2 (p_3+p_5)^2}  +\nonumber\\
	       +&& \frac{i \kappa^3 (\e_1.p_2)^2 (\e_2.\e_4)^2 (\e_3.p_5)^2 (\e_5.p_2) (\e_5.p_4)}{(p_1+p_2)^2 (p_3+p_5)^2} +\nonumber\\+&& \frac{i \kappa^3 (\e_1.p_2)^2 (\e_2.\e_4)^2 (\e_3.p_4) (\e_3.p_5) (\e_5.p_3) (\e_5.p_4)}{(p_1+p_2)^2 (p_3+p_5)^2} -\nonumber\\
	       -&& \frac{i \kappa^3 (\e_1.p_2)^2 (\e_2.\e_4)^2 (\e_3.p_4) (\e_3.p_5) (\e_5.p_2) (\e_5.p_3)}{(p_1+p_2)^2 (p_3+p_5)^2} -\nonumber\\- && \frac{2i \kappa^3 (\e_1.p_2)^2 (\e_2.\e_3) (\e_2.\e_4) (\e_3.p_4) (\e_4.p_2) (\e_5.p_3)^2}{(p_1+p_2)^2 (p_3+p_5)^2}\nonumber\\
	       \eea
	
	       \bea
	       {\cal A}(1^-,2^-;5^+;3^+,4^+)&&=- \frac{i \kappa^3 (\e_1.p_2)^2 (\e_2.\e_4)^2 (\e_3.p_4)^2 (\e_5.p_2)^2}{(p_1+p_2)^2 (p_3+p_4)^2}  - \frac{i \kappa^3 (\e_1.p_2)^2 (\e_2.\e_3)^2 (\e_4.p_3)^2 (\e_5.p_2)^2}{(p_1+p_2)^2 (p_3+p_4)^2}\nonumber\\
	       +&&  \frac{2i \kappa^3 (\e_1.p_2)^2 (\e_2.\e_3) (\e_2.\e_4) (\e_3.p_4) (\e_4.p_3) (\e_5.p_2)^2}{(p_1+p_2)^2 (p_3+p_4)^2}\\\nonumber\vspace{1cm}\\
	       {\cal A}(1^-,3^+;2^-;4^+,5^+)&&=- \frac{i \kappa^3 (\e_1.p_3)^2 (\e_2.\e_4)^2 (\e_3.p_2)^2 (\e_5.p_2)^2}{(p_1+p_3)^2 (p_4+p_5)^2} - \frac{i \kappa^3 (\e_1.p_3)^2 (\e_2.\e_3)^2 (\e_4.p_3)^2 (\e_5.p_2)^2}{(p_1+p_3)^2 (p_4+p_5)^2}-\nonumber\\
	       -&&   \frac{2i \kappa^3 (\e_1.p_3)^2 (\e_2.\e_4)^2 (\e_3.p_2)^2 (\e_5.p_2) (\e_5.p_3)}{(p_1+p_3)^2 (p_4+p_5)^2}-  \frac{i \kappa^3 (\e_1.p_3)^2 (\e_2.\e_3)^2 (\e_4.p_2)^2 (\e_5.p_3)^2}{(p_1+p_3)^2 (p_4+p_5)^2} +\nonumber\\
	       +&& \frac{2i \kappa^3 (\e_1.p_3)^2 (\e_2.\e_3)^2 (\e_4.p_2) (\e_4.p_3) (\e_5.p_2) (\e_5.p_3)}{(p_1+p_3)^2 (p_4+p_5)^2} +\nonumber\\
	       -&&  \frac{2i \kappa^3 (\e_1.p_3)^2 (\e_2.\e_3) (\e_2.\e_4) (\e_3.p_2) (\e_4.p_3) (\e_5.p_2)^2}{(p_1+p_3)^2 (p_4+p_5)^2} -\nonumber\\
	       +&& \frac{2i \kappa^3 (\e_1.p_3)^2 (\e_2.\e_3) (\e_2.\e_4) (\e_3.p_2) (\e_4.p_2) (\e_5.p_2) (\e_5.p_3)}{(p_1+p_3)^2 (p_4+p_5)^2} -\nonumber\\
	       -&&  \frac{2i \kappa^3 (\e_1.p_3)^2 (\e_2.\e_3) (\e_2.\e_4) (\e_3.p_2) (\e_4.p_3) (\e_5.p_2) (\e_5.p_3)}{(p_1+p_3)^2 (p_4+p_5)^2} +\nonumber\\
	       +&& \frac{2i \kappa^3 (\e_1.p_3)^2 (\e_2.\e_3) (\e_2.\e_4) (\e_3.p_2) (\e_4.p_2) (\e_5.p_3)^2}{(p_1+p_3)^2 (p_4+p_5)^2} -\nonumber\\-&&  \frac{i \kappa^3 (\e_1.p_3)^2 (\e_2.\e_4)^2 (\e_3.p_2)^2 (\e_5.p_3)^2}{(p_1+p_3)^2 (p_4+p_5)^2}
	       \\\nonumber\vspace{1cm}\\
	       {\cal A}(1^-,3^+;4^+;2^-,5^+)&&=- \frac{i \kappa^3 (\e_1.p_3)^2 (\e_2.\e_4)^2 (\e_3.p_4)^2 (\e_5.p_2)^2}{(p_1+p_3)^2 (p_2+p_5)^2}- \frac{i \kappa^3 (\e_1.p_3)^2 (\e_2.\e_3)^2 (\e_4.p_3)^2 (\e_5.p_2)^2}{(p_1+p_3)^2 (p_2+p_5)^2}+\nonumber\\+&&   \frac{2i \kappa^3 (\e_1.p_3)^2 (\e_2.\e_3) (\e_2.\e_4) (\e_3.p_4) (\e_4.p_3) (\e_5.p_2)^2}{(p_1+p_3)^2 (p_2+p_5)^2}
	       \\\nonumber\vspace{1cm}\\
	       {\cal A}(1^-,3^+;5^+;2^-,4^+)&&=- \frac{i \kappa^3 (\e_1.p_3)^2 (\e_2.\e_4)^2 (\e_3.p_2)^2 (\e_5.p_2)^2}{(p_1+p_3)^2 (p_2+p_4)^2}-  \frac{i \kappa^3 (\e_1.p_3)^2 (\e_2.\e_3)^2 (\e_4.p_2)^2 (\e_5.p_3)^2}{(p_1+p_3)^2 (p_2+p_4)^2}  -\nonumber\\-&&  \frac{i \kappa^3 (\e_1.p_3)^2 (\e_2.\e_4)^2 (\e_3.p_4)^2 (\e_5.p_2)^2}{(p_1+p_3)^2 (p_2+p_4)^2} -  \frac{2i \kappa^3 (\e_1.p_3)^2 (\e_2.\e_4)^2 (\e_3.p_2)^2 (\e_5.p_2) (\e_5.p_3)}{(p_1+p_3)^2 (p_2+p_4)^2} -\nonumber\\-&&  \frac{2i \kappa^3 (\e_1.p_3)^2 (\e_2.\e_4)^2 (\e_3.p_2) (\e_3.p_4) (\e_5.p_2)^2}{(p_1+p_3)^2 (p_2+p_4)^2}-  \frac{i \kappa^3 (\e_1.p_3)^2 (\e_2.\e_4)^2 (\e_3.p_2)^2 (\e_5.p_3)^2}{(p_1+p_3)^2 (p_2+p_4)^2} -\nonumber\\-&&  \frac{2i \kappa^3 (\e_1.p_3)^2 (\e_2.\e_4)^2 (\e_3.p_2) (\e_3.p_4) (\e_5.p_2) (\e_5.p_3)}{(p_1+p_3)^2 (p_2+p_4)^2} +\nonumber\\+&& \frac{2i \kappa^3 (\e_1.p_3)^2 (\e_2.\e_3) (\e_2.\e_4) (\e_3.p_2) (\e_4.p_2) (\e_5.p_2) (\e_5.p_3)}{(p_1+p_3)^2 (p_2+p_4)^2} +\nonumber\\+&& \frac{2i \kappa^3 (\e_1.p_3)^2 (\e_2.\e_3) (\e_2.\e_4) (\e_3.p_4) (\e_4.p_2) (\e_5.p_2) (\e_5.p_3)}{(p_1+p_3)^2 (p_2+p_4)^2} +\nonumber\\+&& \frac{2i \kappa^3 (\e_1.p_3)^2 (\e_2.\e_3) (\e_2.\e_4) (\e_3.p_2) (\e_4.p_2) (\e_5.p_3)^2}{(p_1+p_3)^2 (p_2+p_4)^2}
	       \eea
	
	       \bea
	       {\cal A}(1^-,4^+;2^-;3^+,5^+)&&=- \frac{i \kappa^3 (\e_1.p_4)^2 (\e_2.\e_4)^2 (\e_3.p_2)^2 (\e_5.p_2)^2}{(p_1+p_4)^2 (p_3+p_5)^2} -  \frac{2i \kappa^3 (\e_1.p_4)^2 (\e_2.\e_4)^2 (\e_3.p_2) (\e_3.p_4) (\e_5.p_2)^2}{(p_1+p_4)^2 (p_3+p_5)^2} - \nonumber\\-&& \frac{i \kappa^3 (\e_1.p_4)^2 (\e_2.\e_4)^2 (\e_3.p_4)^2 (\e_5.p_2)^2}{(p_1+p_4)^2 (p_3+p_5)^2} -  \frac{2i \kappa^3 (\e_1.p_4)^2 (\e_2.\e_4)^2 (\e_3.p_2)^2 (\e_5.p_2) (\e_5.p_3)}{(p_1+p_4)^2 (p_3+p_5)^2}\nonumber\\-&&  \frac{i \kappa^3 (\e_1.p_4)^2 (\e_2.\e_3)^2 (\e_4.p_2)^2 (\e_5.p_3)^2}{(p_1+p_4)^2 (p_3+p_5)^2}-  \frac{i \kappa^3 (\e_1.p_4)^2 (\e_2.\e_4)^2 (\e_3.p_2)^2 (\e_5.p_3)^2}{(p_1+p_4)^2 (p_3+p_5)^2} -\nonumber\\-&&  \frac{2i \kappa^3 (\e_1.p_4)^2 (\e_2.\e_4)^2 (\e_3.p_2) (\e_3.p_4) (\e_5.p_2) (\e_5.p_3)}{(p_1+p_4)^2 (p_3+p_5)^2} +\nonumber\\+&& \frac{2i \kappa^3 (\e_1.p_4)^2 (\e_2.\e_3) (\e_2.\e_4) (\e_3.p_2) (\e_4.p_2) (\e_5.p_2) (\e_5.p_3)}{(p_1+p_4)^2 (p_3+p_5)^2} +\nonumber\\+&& \frac{2i \kappa^3 (\e_1.p_4)^2 (\e_2.\e_3) (\e_2.\e_4) (\e_3.p_4) (\e_4.p_2) (\e_5.p_2) (\e_5.p_3)}{(p_1+p_4)^2 (p_3+p_5)^2}  +\nonumber\\+&& \frac{2i \kappa^3 (\e_1.p_4)^2 (\e_2.\e_3) (\e_2.\e_4) (\e_3.p_2) (\e_4.p_2) (\e_5.p_3)^2}{(p_1+p_4)^2 (p_3+p_5)^2}
	       \eea

	       \bea
	       {\cal A}(1^-,4^+;3^+;2^-,5^+)&&= - \frac{i \kappa^3 \left(e_1.t_4\right)^2 \left(e_2.e_4\right)^2 \left(e_3.p_4\right)^2 \left(e_5.t_2\right)^2}{q^2 p^2}-  \frac{i \kappa^3 \left(e_1.t_4\right)^2 \left(e_2.e_3\right)^2 \left(e_4.t_3\right)^2 \left(e_5.t_2\right)^2}{q^2 p^2} +\nonumber\\+&& \frac{2i \kappa^3 \left(e_1.t_4\right)^2 \left(e_2.e_3\right) \left(e_2.e_4\right) \left(e_3.p_4\right) \left(e_4.t_3\right) \left(e_5.t_2\right)^2}{q^2 p^2} 
	       \eea
	
	       \bea
	       {\cal A}(1^-,4^+;5^+;2^-,3^+)&&=- \frac{i \kappa^3 (\e_1.p_4)^2 (\e_2.\e_4)^2 (\e_3.p_2)^2 (\e_5.p_2)^2}{(p_1+p_4)^2 (p_2+p_3)^2}-  \frac{i \kappa^3 (\e_1.p_4)^2 (\e_2.\e_3)^2 (\e_4.p_2)^2 (\e_5.p_3)^2}{(p_1+p_4)^2 (p_2+p_3)^2}-\nonumber\\-&&  \frac{i \kappa^3 (\e_1.p_4)^2 (\e_2.\e_3)^2 (\e_4.p_3)^2 (\e_5.p_2)^2}{(p_1+p_4)^2 (p_2+p_3)^2} -  \frac{i \kappa^3 (\e_1.p_4)^2 (\e_2.\e_4)^2 (\e_3.p_2)^2 (\e_5.p_3)^2}{(p_1+p_4)^2 (p_2+p_3)^2}-\nonumber\\ -&&  \frac{2i \kappa^3 (\e_1.p_4)^2 (\e_2.\e_3) (\e_2.\e_4) (\e_3.p_2) (\e_4.p_3) (\e_5.p_2)^2}{(p_1+p_4)^2 (p_2+p_3)^2} -\nonumber\\-&&  \frac{2i \kappa^3 (\e_1.p_4)^2 (\e_2.\e_4)^2 (\e_3.p_2)^2 (\e_5.p_2) (\e_5.p_3)}{(p_1+p_4)^2 (p_2+p_3)^2} +\nonumber\\+&& \frac{2i \kappa^3 (\e_1.p_4)^2 (\e_2.\e_3) (\e_2.\e_4) (\e_3.p_2) (\e_4.p_2) (\e_5.p_2) (\e_5.p_3)}{(p_1+p_4)^2 (p_2+p_3)^2} -\nonumber\\-&&  \frac{2i \kappa^3 (\e_1.p_4)^2 (\e_2.\e_3) (\e_2.\e_4) (\e_3.p_2) (\e_4.p_3) (\e_5.p_2) (\e_5.p_3)}{(p_1+p_4)^2 (p_2+p_3)^2} +\nonumber\\+&& \frac{2i \kappa^3 (\e_1.p_4)^2 (\e_2.\e_3)^2 (\e_4.p_2) (\e_4.p_3) (\e_5.p_2) (\e_5.p_3)}{(p_1+p_4)^2 (p_2+p_3)^2}  +\nonumber\\+&& \frac{2i \kappa^3 (\e_1.p_4)^2 (\e_2.\e_3) (\e_2.\e_4) (\e_3.p_2) (\e_4.p_2) (\e_5.p_3)^2}{(p_1+p_4)^2 (p_2+p_3)^2}\\
	       {\cal A}(1^-,5^+;2^-;3^+,4^+)&&=0\\
	       \nonumber\\
	       {\cal A}(1^-,5^+;3^+;2^-,4^+)&&=0\\
	       \nonumber\\
	       {\cal A}(1^-,5^+;4^+;2^-,3^+)&&=0
	       \eea
	
	       \bea
	       {\cal A}(2^-,3^+;1^-;4^+,5^+)&&=\frac{i \kappa^3 (\e_1.p_2) (\e_1.p_4) (\e_2.\e_4)^2 (\e_3.p_2)^2 (\e_5.p_2)^2}{(p_2+p_3)^2 (p_4+p_5)^2} +\nonumber\\+&& \frac{i \kappa^3 (\e_1.p_3) (\e_1.p_4) (\e_2.\e_4)^2 (\e_3.p_2)^2 (\e_5.p_2)^2}{(p_2+p_3)^2 (p_4+p_5)^2} +\nonumber \\+&& \frac{2i \kappa^3 (\e_1.p_2) (\e_1.p_4) (\e_2.\e_3) (\e_2.\e_4) (\e_3.p_2) (\e_4.p_3) (\e_5.p_2)^2}{(p_2+p_3)^2 (p_4+p_5)^2} +\nonumber \\+&& \frac{2i \kappa^3 (\e_1.p_3) (\e_1.p_4) (\e_2.\e_3) (\e_2.\e_4) (\e_3.p_2) (\e_4.p_3) (\e_5.p_2)^2}{(p_2+p_3)^2 (p_4+p_5)^2} +\nonumber \\+&& \frac{i \kappa^3 (\e_1.p_2) (\e_1.p_4) (\e_2.\e_3)^2 (\e_4.p_3)^2 (\e_5.p_2)^2}{(p_2+p_3)^2 (p_4+p_5)^2} +\nonumber\\
	       +&& \frac{i \kappa^3 (\e_1.p_3) (\e_1.p_4) (\e_2.\e_3)^2 (\e_4.p_3)^2 (\e_5.p_2)^2}{(p_2+p_3)^2 (p_4+p_5)^2} +\nonumber \\+&& \frac{2i \kappa^3 (\e_1.p_2) (\e_1.p_4) (\e_2.\e_4)^2 (\e_3.p_2)^2 (\e_5.p_2) (\e_5.p_3)}{(p_2+p_3)^2 (p_4+p_5)^2} +\nonumber \\+&& \frac{2i \kappa^3 (\e_1.p_3) (\e_1.p_4) (\e_2.\e_4)^2 (\e_3.p_2)^2 (\e_5.p_2) (\e_5.p_3)}{(p_2+p_3)^2 (p_4+p_5)^2} -\nonumber \\-&&  \frac{2i \kappa^3 (\e_1.p_2) (\e_1.p_4) (\e_2.\e_3) (\e_2.\e_4) (\e_3.p_2) (\e_4.p_2) (\e_5.p_2) (\e_5.p_3)}{(p_2+p_3)^2 (p_4+p_5)^2} -\nonumber \\-&&  \frac{2i \kappa^3 (\e_1.p_3) (\e_1.p_4) (\e_2.\e_3) (\e_2.\e_4) (\e_3.p_2) (\e_4.p_2) (\e_5.p_2) (\e_5.p_3)}{(p_2+p_3)^2 (p_4+p_5)^2} +\nonumber \\+&& \frac{2i \kappa^3 (\e_1.p_2) (\e_1.p_4) (\e_2.\e_3) (\e_2.\e_4) (\e_3.p_2) (\e_4.p_3) (\e_5.p_2) (\e_5.p_3)}{(p_2+p_3)^2 (p_4+p_5)^2} +\nonumber \\+&& \frac{2i \kappa^3 (\e_1.p_3) (\e_1.p_4) (\e_2.\e_3) (\e_2.\e_4) (\e_3.p_2) (\e_4.p_3) (\e_5.p_2) (\e_5.p_3)}{(p_2+p_3)^2 (p_4+p_5)^2} -\nonumber \\-&&  \frac{2i \kappa^3 (\e_1.p_2) (\e_1.p_4) (\e_2.\e_3)^2 (\e_4.p_2) (\e_4.p_3) (\e_5.p_2) (\e_5.p_3)}{(p_2+p_3)^2 (p_4+p_5)^2} -\nonumber \\-&&  \frac{2i \kappa^3 (\e_1.p_3) (\e_1.p_4) (\e_2.\e_3)^2 (\e_4.p_2) (\e_4.p_3) (\e_5.p_2) (\e_5.p_3)}{(p_2+p_3)^2 (p_4+p_5)^2} +\nonumber \\+&& \frac{i \kappa^3 (\e_1.p_2) (\e_1.p_4) (\e_2.\e_4)^2 (\e_3.p_2)^2 (\e_5.p_3)^2}{(p_2+p_3)^2 (p_4+p_5)^2} +\nonumber\\
	       +&& \frac{i \kappa^3 (\e_1.p_3) (\e_1.p_4) (\e_2.\e_4)^2 (\e_3.p_2)^2 (\e_5.p_3)^2}{(p_2+p_3)^2 (p_4+p_5)^2} -\nonumber \\-&&  \frac{2i \kappa^3 (\e_1.p_2) (\e_1.p_4) (\e_2.\e_3) (\e_2.\e_4) (\e_3.p_2) (\e_4.p_2) (\e_5.p_3)^2}{(p_2+p_3)^2 (p_4+p_5)^2} -\nonumber \\-&&  \frac{2i \kappa^3 (\e_1.p_3) (\e_1.p_4) (\e_2.\e_3) (\e_2.\e_4) (\e_3.p_2) (\e_4.p_2) (\e_5.p_3)^2}{(p_2+p_3)^2 (p_4+p_5)^2} +\nonumber \\+&&\frac{i \kappa^3 (\e_1.p_2) (\e_1.p_4) (\e_2.\e_3)^2 (\e_4.p_2)^2 (\e_5.p_3)^2}{(p_2+p_3)^2 (p_4+p_5)^2} +\nonumber\\
	       +&& \frac{i \kappa^3 (\e_1.p_3) (\e_1.p_4) (\e_2.\e_3)^2 (\e_4.p_2)^2 (\e_5.p_3)^2}{q^2 (p_4+p_5)^2}
	       \eea
	
	       \bea
	       {\cal A}(2^-,4^+;1^-;3^+,5^+)&&=- \frac{i \kappa^3 (\e_1.p_3)^2 (\e_2.\e_4)^2 (\e_3.p_2)^2 (\e_5.p_2)^2}{(p_2+p_4)^2 (p_3+p_5)^2}-  \frac{i \kappa^3 (\e_1.p_3)^2 (\e_2.\e_3)^2 (\e_4.p_2)^2 (\e_5.p_3)^2}{(p_2+p_4)^2 (p_3+p_5)^2}  -\nonumber \\-&&  \frac{i \kappa^3 (\e_1.p_3)^2 (\e_2.\e_4)^2 (\e_3.p_4)^2 (\e_5.p_2)^2}{(p_2+p_4)^2 (p_3+p_5)^2} -  \frac{i \kappa^3 (\e_1.p_3)^2 (\e_2.\e_4)^2 (\e_3.p_2)^2 (\e_5.p_3)^2}{(p_2+p_4)^2 (p_3+p_5)^2} -\nonumber \\-&&  \frac{2i \kappa^3 (\e_1.p_3)^2 (\e_2.\e_4)^2 (\e_3.p_2) (\e_3.p_4) (\e_5.p_2) (\e_5.p_3)}{(p_2+p_4)^2 (p_3+p_5)^2} +\nonumber \\+&& \frac{2i \kappa^3 (\e_1.p_3)^2 (\e_2.\e_3) (\e_2.\e_4) (\e_3.p_2) (\e_4.p_2) (\e_5.p_2) (\e_5.p_3)}{(p_2+p_4)^2 (p_3+p_5)^2} +\nonumber \\+&& \frac{2i \kappa^3 (\e_1.p_3)^2 (\e_2.\e_3) (\e_2.\e_4) (\e_3.p_4) (\e_4.p_2) (\e_5.p_2) (\e_5.p_3)}{(p_2+p_4)^2 (p_3+p_5)^2} +\nonumber \\+&& \frac{2i \kappa^3 (\e_1.p_3)^2 (\e_2.\e_3) (\e_2.\e_4) (\e_3.p_2) (\e_4.p_2) (\e_5.p_3)^2}{(p_2+p_4)^2 (p_3+p_5)^2} -\nonumber\\-&& \frac{2i \kappa^3 (\e_1.p_3)^2 (\e_2.\e_4)^2 (\e_3.p_2) (\e_3.p_4) (\e_5.p_2)^2}{(p_2+p_4)^2 (p_3+p_5)^2}\nonumber-\\ -&&  \frac{2i \kappa^3 (\e_1.p_3)^2 (\e_2.\e_4)^2 (\e_3.p_2)^2 (\e_5.p_2) (\e_5.p_3)}{(p_2+p_4)^2 (p_3+p_5)^2}     \\\nonumber\vspace{1cm}\\
	       {\cal A}(2^-,4^+;1^-;3^+,5^+)&&=\frac{i \kappa^3 (\e_1.p_2) (\e_1.p_3) (\e_2.\e_4)^2 (\e_3.p_4)^2 (\e_5.p_2)^2}{(p_2+p_5)^2 (p_3+p_4)^2} +\nonumber\\+&& \frac{i \kappa^3 (\e_1.p_2) (\e_1.p_4) (\e_2.\e_4)^2 (\e_3.p_4)^2 (\e_5.p_2)^2}{(p_2+p_5)^2 (p_3+p_4)^2} +\nonumber \\+&& \frac{i \kappa^3 (\e_1.p_2) (\e_1.p_3) (\e_2.\e_3)^2 (\e_4.p_3)^2 (\e_5.p_2)^2}{(p_2+p_5)^2 (p_3+p_4)^2} +\nonumber\\+&& \frac{i \kappa^3 (\e_1.p_2) (\e_1.p_4) (\e_2.\e_3)^2 (\e_4.p_3)^2 (\e_5.p_2)^2}{(p_2+p_5)^2 (p_3+p_4)^2}-\nonumber \\
	       -&&  \frac{2i \kappa^3 (\e_1.p_2) (\e_1.p_3) (\e_2.\e_3) (\e_2.\e_4) (\e_3.p_4) (\e_4.p_3) (\e_5.p_2)^2}{(p_2+p_5)^2 (p_3+p_4)^2} -\nonumber \\-&&  \frac{2i \kappa^3 (\e_1.p_2) (\e_1.p_4) (\e_2.\e_3) (\e_2.\e_4) (\e_3.p_4) (\e_4.p_3) (\e_5.p_2)^2}{(p_2+p_5)^2 (p_3+p_4)^2}
	       \eea


\newpage

     \begin{thebibliography}{20}


	\bibitem{AlvarezGMHVM}
	E.~Alvarez, S.~Gonzalez-Martin, M.~Herrero-Valea and C.~P.~Martin,
	``Quantum Corrections to Unimodular Gravity,''
	JHEP {\bf 1508} (2015) 078
	doi:10.1007/JHEP08(2015)078
	[arXiv:1505.01995 [hep-th]].
\bibitem{Redux}
  E.~Alvarez, S.~Gonzalez-Martin, M.~Herrero-Valea and C.~P.~Martin,
  ``Unimodular Gravity Redux,''
  Phys.\ Rev.\ D {\bf 92} (2015) no.6,  061502
  doi:10.1103/PhysRevD.92.061502
  [arXiv:1505.00022 [hep-th]].


     	
     
\bibitem{Elvang:2015rqa}
  H.~Elvang and Y.~t.~Huang,
``Scattering Amplitudes in Gauge Theory and Gravity,''
Cambridge University Press (2015-04-02),ISBN: 9781316191422 (eBook), 9781107069251 (Print)

\bibitem{Benincasa}
P.~Benincasa, C.~Boucher-Veronneau and F.~Cachazo,
``Taming Tree Amplitudes In General Relativity,''
JHEP {\bf 0711} (2007) 057
doi:10.1088/1126-6708/2007/11/057
[hep-th/0702032 [HEP-TH]].\\
\bibitem{Ananth}
  S.~Ananth and S.~Theisen,
  ``KLT relations from the Einstein-Hilbert Lagrangian,''
  Phys.\ Lett.\ B {\bf 652} (2007) 128
  doi:10.1016/j.physletb.2007.07.003
  [arXiv:0706.1778 [hep-th]].
  \bibitem{Berends}
  F.~A.~Berends, W.~T.~Giele and H.~Kuijf,
  ``On relations between multi - gluon and multigraviton scattering,''
  Phys.\ Lett.\ B {\bf 211} (1988) 91.
  doi:10.1016/0370-2693(88)90813-1
  
  
\bibitem{Cachazo}
  F.~Cachazo and P.~Svrcek,
  ``Tree level recursion relations in general relativity,''
  hep-th/0502160.
  

    \bibitem{Benincasa2}
    P.~Benincasa and F.~Cachazo,
    ``Consistency Conditions on the S-Matrix of Massless Particles,''
    arXiv:0705.4305 [hep-th].

\bibitem{Henn}
J.~M.~Henn and J.~C.~Plefka,
``Scattering Amplitudes in Gauge Theories,''
Lect.\ Notes Phys.\  {\bf 883} (2014) 1.
doi:978-3-642-54021-9, 10.1007/978-3-642-54022-6

\bibitem{Berends:1987cv}
  F.~A.~Berends and W.~Giele,
  Nucl.\ Phys.\ B {\bf 294} (1987) 700.
  doi:10.1016/0550-3213(87)90604-3




 N.~Arkani-Hamed and J.~Kaplan,
  ``On Tree Amplitudes in Gauge Theory and Gravity,''
  JHEP {\bf 0804} (2008) 076
  doi:10.1088/1126-6708/2008/04/076
  [arXiv:0801.2385 [hep-th]].\\
 N.~E.~J.~Bjerrum-Bohr, P.~H.~Damgaard, B.~Feng and T.~Sondergaard,
  ``Proof of Gravity and Yang-Mills Amplitude Relations,''
  JHEP {\bf 1009} (2010) 067
  doi:10.1007/JHEP09(2010)067
  [arXiv:1007.3111 [hep-th]].\\
  H.~Elvang and D.~Z.~Freedman,
  ``Note on graviton MHV amplitudes,''
  JHEP {\bf 0805}, 096 (2008)
  doi:10.1088/1126-6708/2008/05/096
  [arXiv:0710.1270 [hep-th]].

    N.~E.~J.~Bjerrum-Bohr, D.~C.~Dunbar, H.~Ita, W.~B.~Perkins and K.~Risager,
    ``MHV-vertices for gravity amplitudes,''
    JHEP {\bf 0601} (2006) 009
    doi:10.1088/1126-6708/2006/01/009
    [hep-th/0509016].

    J.~Bedford, A.~Brandhuber, B.~J.~Spence and G.~Travaglini,
    ``A Recursion relation for gravity amplitudes,''
    Nucl.\ Phys.\ B {\bf 721} (2005) 98
    doi:10.1016/j.nuclphysb.2005.016
    [hep-th/0502146].













     	\bibitem{Goroff}
     	M.~H.~Goroff and A.~Sagnotti,
     	``The Ultraviolet Behavior of Einstein Gravity,''
     	Nucl.\ Phys.\ B {\bf 266} (1986) 709.
     	doi:10.1016/0550-3213(86)90193-8
     	
     	\bibitem{Sannan}
     	S.~Sannan,
     	``Gravity as the Limit of the Type {II} Superstring Theory,''
     	Phys.\ Rev.\ D {\bf 34} (1986) 1749.
     	doi:10.1103/PhysRevD.34.1749
     	
     	\bibitem{DeWitt}
     	B.~S.~DeWitt,
     	``Quantum Theory of Gravity. 3. Applications of the Covariant Theory,''
     	Phys.\ Rev.\  {\bf 162} (1967) 1239.
     	doi:10.1103/PhysRev.162.1239
\bibitem{Britto}
  R.~Britto, F.~Cachazo, B.~Feng and E.~Witten,
  ``Direct proof of tree-level recursion relation in Yang-Mills theory,''
  Phys.\ Rev.\ Lett.\  {\bf 94} (2005) 181602
  doi:10.1103/PhysRevLett.94.181602
  [hep-th/0501052].


\bibitem{Kuipers:2012rf}
  J.~Kuipers, T.~Ueda, J.~A.~M.~Vermaseren and J.~Vollinga,
  ``FORM version 4.0,''
  Comput.\ Phys.\ Commun.\  {\bf 184} (2013) 1453
  doi:10.1016/j.cpc.2012.12.028
  [arXiv:1203.6543 [cs.SC]].


 \bibitem{Martingarcia}
  J.~ M.~ Martín-García et.al. xAct: Efficient tensor computer algebra for Mathematica.2002-2013.url:http://xact.es/
  
  
\bibitem{Donoghue:1994dn}
  J.~F.~Donoghue,
  Phys.\ Rev.\ D {\bf 50} (1994) 3874
  doi:10.1103/PhysRevD.50.3874
  [gr-qc/9405057].
    	
     	\end{thebibliography}
	    \end{document}